\documentclass{PoS}

\pdfoutput=1

\title{The case of 3C326: VLA 74 MHz observations during a geomagnetic storm}

\ShortTitle{The case of 3C326: VLA 74 MHz observations during a geomagnetic storm}

\author{{Emanuela Orr\'u}\\
        Radboud University Nijmegen\\
        Heijendaalseweg 135,   6525 AJ Nijmegen The Netherlands\\
        E-mail: \email{e.orru@astro.ru.nl}}

\author{{Huib Intema}\\
        National Radio Astronomy Observatory\\
        520 Edgemont Road Charlottesville, VA 22903-2475, USA\\
        E-mail: \email{hintema@nrao.edu}}

\abstract{Reaching the thermal noise at low frequencies with the next
  generation of instruments (e.g. SKA, LOFAR etc.) is going to be a challenge. It requires the development of more advanced techniques of calibration compared to those used from the traditional radio astronomy until now. This revolution has slowly started, from self-cal, going through field based correction and SPAM up to the formulation and application of a general Measurement Equation. 
We will describe and compare the several approaches of calibration used so far to reduce low frequency data. 
We will present some results of a 74 MHz VLA observation in exceptional ionospheric conditions of the giant radio galaxy 3C326 for which some of these methods have been successfully applied.}

\FullConference{ISKAF2010 Science Meeting - ISKAF2010\\
		June 10-14, 2010\\
		Assen, the Netherlands}

\begin{document}
\section{Introduction}

\begin{figure}
\includegraphics[width=.3\textwidth]{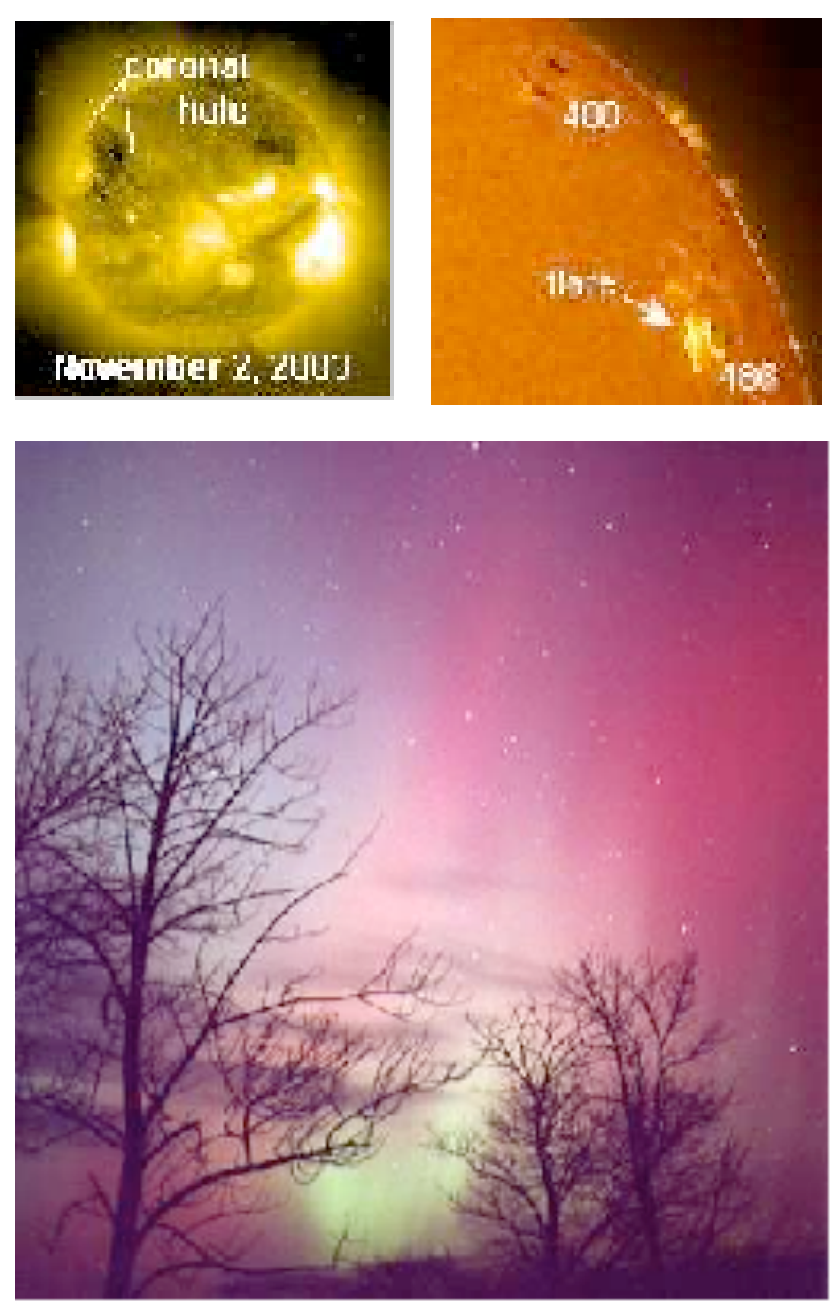}
\caption{(Letf) Coronal holes, (Right) sunspot and
  aurora pictures (center). From spaceweather.com ``Another remarkable solar flare has erupted
  on Nov. 2nd 2003. As a result of the flare, solar protons are
  streaming past Earth. The ongoing radiation storm is a strong
  S3-class event. Passengers and crew in commercial jets at high
  latitudes may receive low-level radiation exposure
  approximately equal to one chest x-ray.''}
\label{fig1}
\end{figure}

\begin{figure}
\includegraphics[width=.6\textwidth]{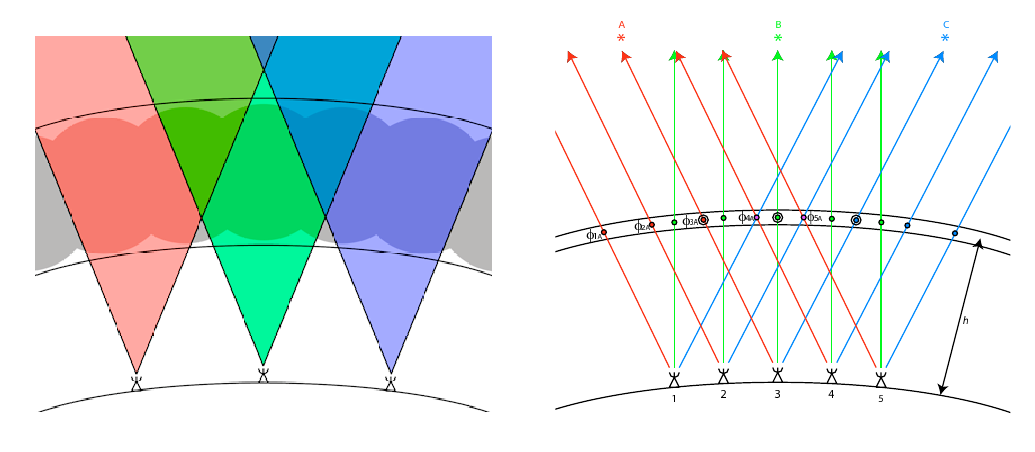}
\caption{ {\bf a)} The incident extraterrestrial signals, received on each
  ground-based antenna, are affected by the ionosphere (grey blobs)
  within cone-like volumes (red, green, blue) that extends over the
  angular width of the field of view. The most dominant effects on
  radio astronomical observations are the ionosphere-induced
  differential propagation delays and Faraday rotations, effects that
  vary with time, antenna-position and viewing direction. {\bf b)} Using calibration results towards several bright sources
  (red, green, blue) in the field of view, SPAM maps the calibration
  phases onto a fixed layer at the ionospheric pierce point positions
  (IPPs) of the lines-of-sight from antennas to sources. For each
  calibration time, SPAM fits an overall phase screen to these IPPs,
  using an optimized set of orthogonal base functions (van der Tol \&
  van der Veen 2007). This model is used to interpolate the
  calibration phases to new IPPs along arbitrary lines-of-sight.}
\label{fig2}
\end{figure}

Spatial and temporal variations in the ionospheric free electron
density can severely effect the delicate astronomical radio
interferometric obsevations. One dominant effect is a propagation
delay that depends on the ray path through the ionosphere, and
therefore depends on time, the position of the interferometer elements
(the antennas or stations) and the viewing direction. The free electron
content of the ionosphere varies with time of day, season,
geographic latitude and Solar activity. Observing with a radio
interferometer in the presence of a thick ionosphere is depicted
schematically in fig.\ref{fig2}a. Here we present a VLA observation at 74 MHz of
the source 3C326 obtained in extreme ionospheric conditions during a
geomagnetic storm (Figure \ref{fig1}) for which the Source Peeling \&
Atmospheric Modeling (SPAM) algorithm has been applied.

\section{The method: SPAM}

Typical interferometric observations at low radio frequencies (<300
MHz) require a calibration algorithm different from self-calibration,
to handle direction-dependent phase errors across the field of view.
Currently, only very few direction-dependent calibration algorithms
exist. The SPAM algorithm (Intema et al. 2009) was developed specifically
for direction-dependent ionospheric calibration and subsequent
imaging. Calibration is done by `peeling' available calibrators in the
field of view (e.g., Nijboer \& Noordam 2007), fitting a time-variant
ionospheric phase screen model to the peeling phase solutions, and
applying the direction-dependent model phase solutions while imaging
the full field of view. For the removal of differential propagation
delays, SPAM uses an ionosphere model that consists of one or several
thin layers. Figure \ref{fig2}b sketches the model set up using one layer. This
approach is successfully tested on several observations from the VLA
at 74 MHz (A\&B-configurations) and the GMRT at 150 MHz. 
Recently, SPAM has undergone several modifications to improve
convergence of the ionosphere model during more severe ionospheric
conditions. These modifications include a multi-layer ionosphere
model, rejection of systematically bad antennas and calibrator
sources, and an FFT-based estimation of the large-scale phase
gradient. This version of SPAM was used to calibrate and image the
severely distorted observations of 3C326 with the VLA at 74 MHz.

\section{Ionospheric model}

\begin{figure}
\includegraphics[width=.6\textwidth]{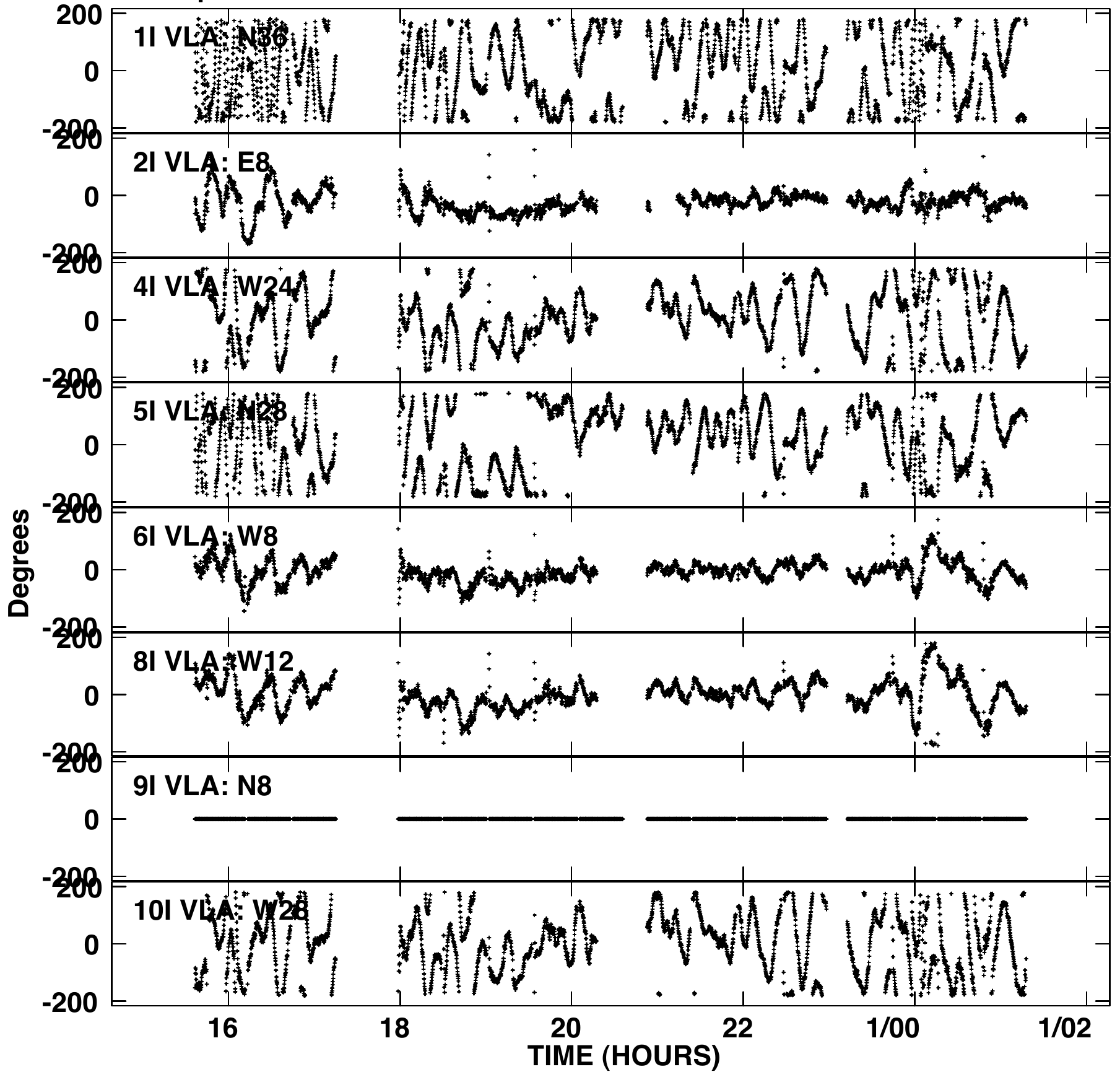}
\caption{Phase solutions obtained from self-calibration.}
\label{fig3}
\end{figure}

As shown in fig. \ref{fig3}, the first part of the observation is characterized
by fast phase variations, evident effect of the extreme conditions of
the ionosphere. Some time ranges were too difficult to model, and were rejected during SPAM calibration.

\begin{figure}
\includegraphics[width=.6\textwidth]{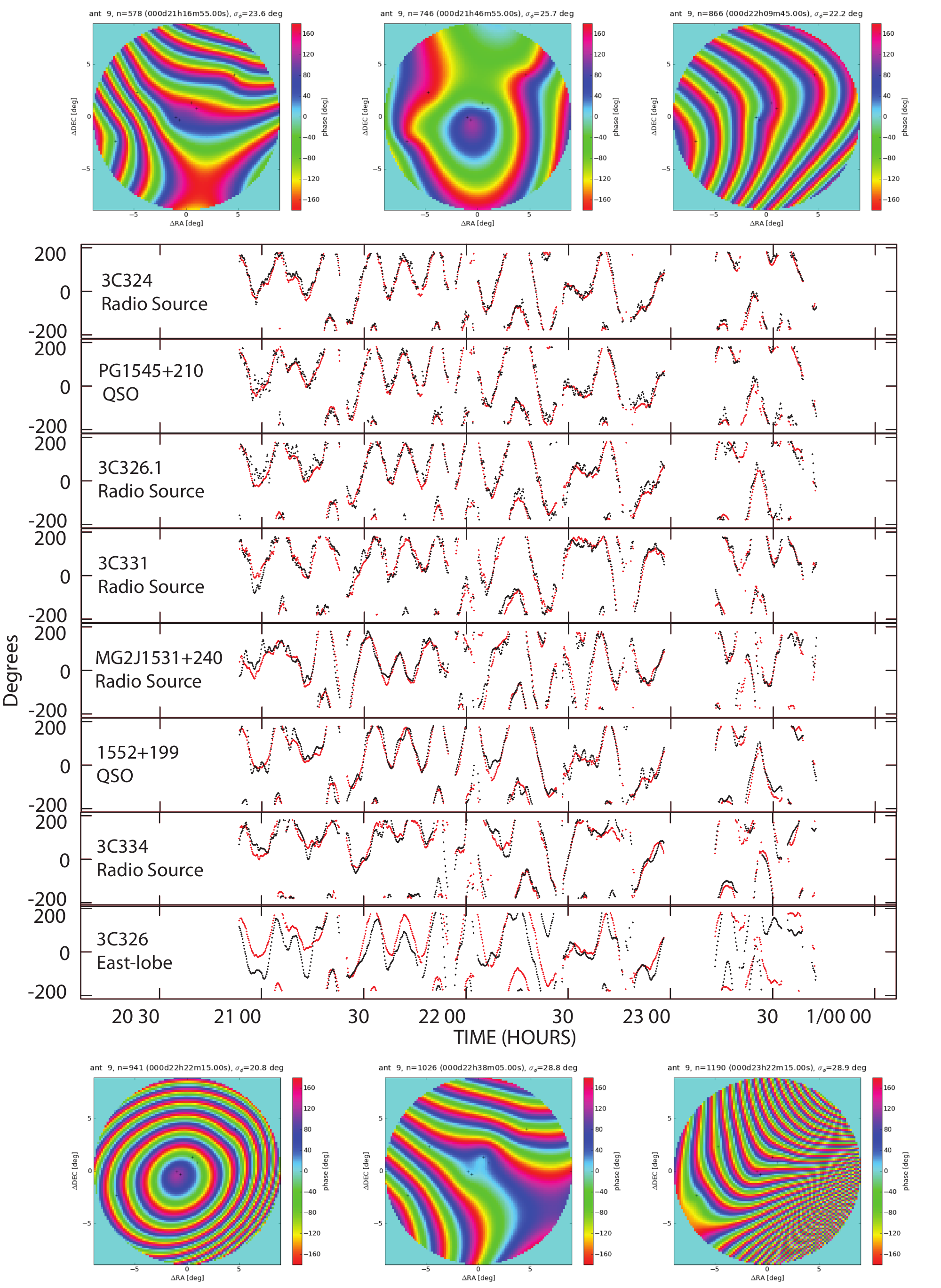}
\caption{See text in ionoshperic model.}
\label{fig4}
\end{figure}

For the peeling process we used eight bright sources found in a radius
of 7.5 degrees. Thus we performed a directional calibration in 8
different patches of the field of view surrounding these sources. The
central panel of fig. \ref{fig4} shows the phase solutions (black dots) of the
eight bright sources used for the peeling fitted with a ionospheric
model (red dots) obtained by the SPAM algorithm. The model fits very
well with the data for the first seven sources, while it seems to miss
some features for the last one, which is the eastern lobe of
3C326. The reason of this discrepancy could be a mismatch in the
assumed and true peak position of this resolved source. Note that the
peeling phase solutions towards this source were excluded from the
ionospheric model fit for exactly this reason. The colored panels on
top and bottom of fig. 4 represent few sketches of the movie which
describes the ionospheric phase variation as fitted by the model. The
time resolution of each frame is 10 seconds, which corresponds to the
integration time of the observation. These plots clearly show how the
ionospherical phases change suddenly passing from very fine and
regular patterns, in principle easy to predict and describe with
models, to wide and irregular features which complicate the
calibration approach.

\begin{figure}
\includegraphics[width=1.0\textwidth]{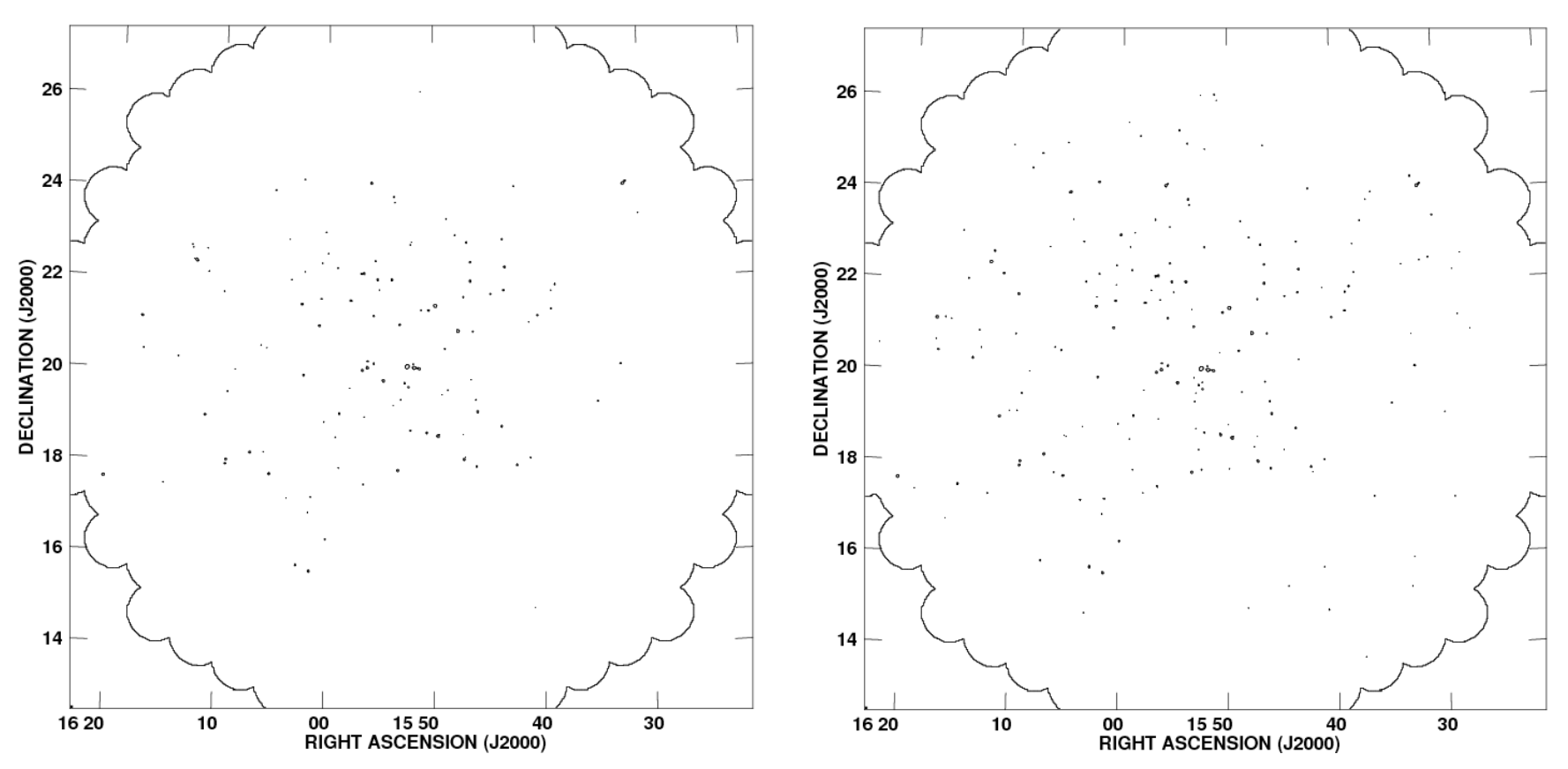}
\caption{Contours of the field of view obtained with self-calibration, (Right) contours of the field of view obtained with SPAM.}
\label{fig5}
\end{figure}

\begin{figure}
\includegraphics[width=.6\textwidth]{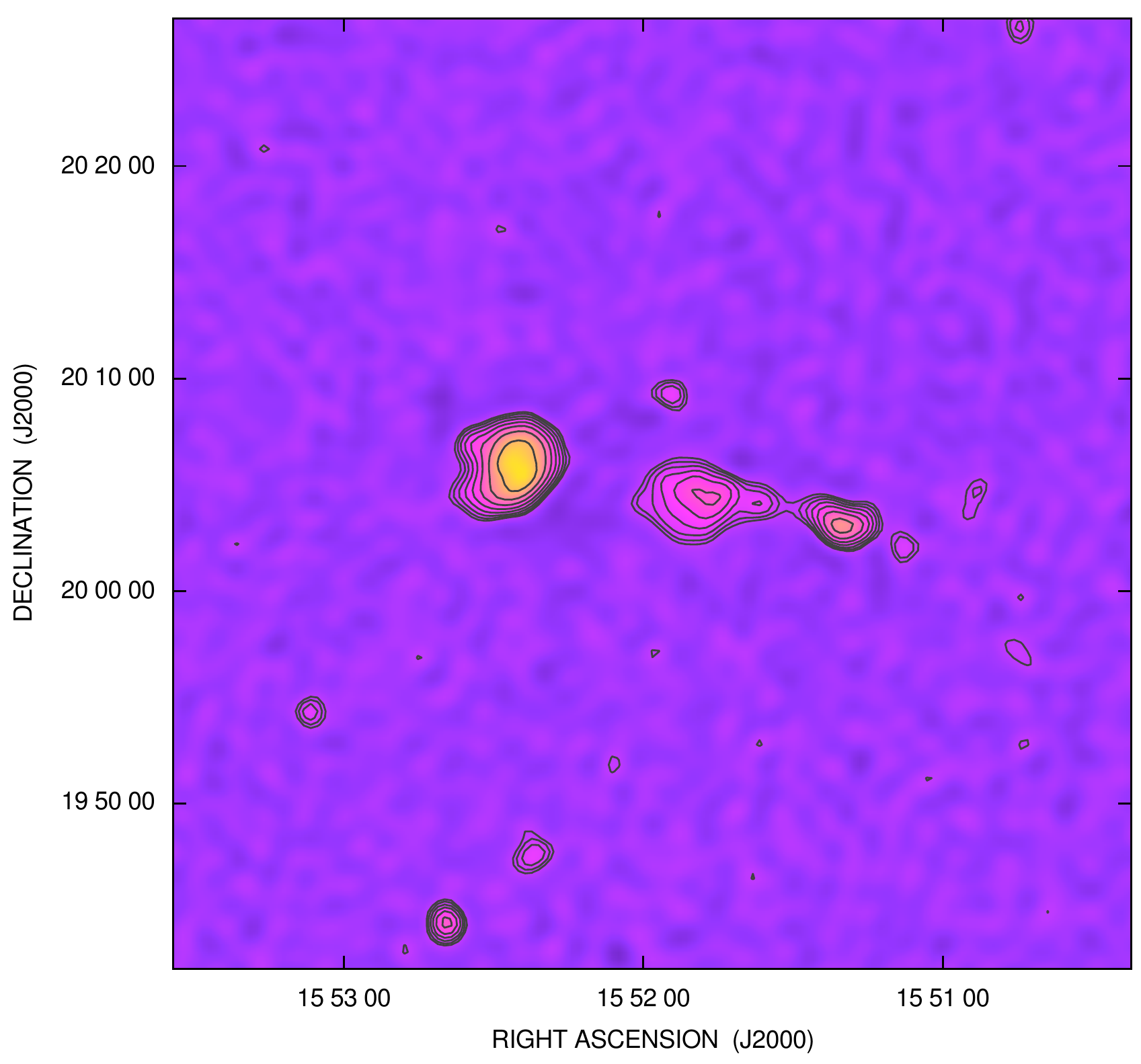}
\caption{VLA image at 74 MHz of the giant radio source 3C326 obtained
  using the SPAM calibration method. Colors and countours show the
  radio brightness at 74 MHz. The rms is $\sim$70 mJy/beam. Contours
  start at 3$\sigma$ level and scale with $\sqrt{2}$. The resolution is
  75$^{\prime\prime}$ $\times$ 75$^{\prime\prime}$.}
\label{fig6}
\end{figure}

\section{SPAM versus self-calibration}
Self-calibration solves for time-variable antenna phases (and
amplitudes) that are assumed to be constant over the
field of view. For low-frequency radio observations, the angular scale
size of ionosphere-induced phase structure over each antenna is
typically smaller than the field of view, thereby degrading the
performance of self-calibration. In those cases, self-calibration has
the tendency to find phase solutions in the direction of the field
source(s) with the highest apparent flux. At larger distance from the
bright source(s), the phase errors increase and source flux will be
more and more scattered. This effect is demonstrated in Figure \ref{fig5},
which shows contour plots of the apparent field of view around 3C326
(no primary beam correction applied). The left plot and right plot
were generated using self-calibration and SPAM, respectively. In
comparison, the source density in the self-calibration plot drops much
faster when moving away from the field center. Note that also the SPAM
plot has several areas near the edges of the field that are relatively
empty, like the SW region. For the latter, this could be the effect of
a lack of bright-enough calibrators in that region to constrain the
ionosphere model more accurately. Figure \ref{fig6} shows the VLA image at
74 MHz of the source 3C326 obtained applying the SPAM algorithm.

\end{document}